\documentclass{aa}
\usepackage{graphicx}
 \usepackage{amsmath}
 \usepackage{amsfonts}
 \usepackage{amssymb}
\usepackage{natbib}

\newcommand{\av}[1]{\langle{#1}\rangle}

\begin{document}


\title{Conceptual problems in detecting the evolution of dark energy
when using distance measurements}

\author{Krzysztof Bolejko}

\institute{Nicolaus Copernicus Astronomical Center,
Polish Academy of Sciences, Bartycka 18, 00 716 Warszawa, Poland \\
\email{bolejko@camk.edu.pl} }

\date{Received ...  ; accepted ...  }

\abstract
{Dark energy is now one of the most important and topical problems in cosmology. The first step to reveal its nature is to detect the evolution of dark energy or to prove beyond doubt that the cosmological constant is indeed constant. However, in the standard approach to cosmology, the Universe is described by the homogeneous and isotropic Friedmann models.}
{We aim to show
that in the perturbed universe (even if perturbations vanish if averaged over sufficiently large scales) the distance-redshift relation is not the same as in the unperturbed universe. This has a serious consequence when studying the nature of dark energy and, as shown here, 
can impair the analysis and studies of dark energy.}
{
The analysis is based on two methods: the linear lensing approximation and 
the non-linear Szekeres Swiss-Cheese model. 
The inhomogeneity scale is $\sim 50$ Mpc, and both models have the same
density fluctuations along the line of sight.}
{The comparison between linear and non-linear methods shows that non-linear
corrections are not negligible.
When inhomogeneities are present the distance changes by several percent.
To show how this change influences the measurements of dark energy,
ten future observations with 2 \% uncertainties are generated.
It is shown the using the standard methods
(i.e. under the assumption of homogeneity) 
the systematics due to inhomogeneities can distort our analysis,
and may lead to a conclusion that dark energy evolves  when in fact it is constant (or vice versa).
}
{Therefore, if future observations are analysed only within the homogeneous framework then the impact of inhomogeneities (such as voids and superclusters) can be mistaken for evolving dark energy. Since the robust distinction between the evolution and non-evolution of dark energy is the first step  to understanding the nature of dark energy a proper handling of inhomogeneities is essential.}


\authorrunning{K. Bolejko}
\titlerunning{Evolving dark energy and distance measurements}

\maketitle

\section{Introduction}

The most important problem in contemporary cosmology is the nature
of dark energy. 
There are many ongoing attempts to explain the nature of dark energy. The simplest explanation assumes that dark energy does not evolve and is modeled by only one single constant --- the cosmological constant. 
In this scenario, the cosmological constant is considered as one of the fundamental constants of nature. However, most  scientists are not satisfied with this type of solution and 
continue to investigate the nature of dark energy. Most of these attempts invoke new physics: f(R) cosmologies \citep{B70,S80,K82,BO83,BC88,LB07,Ky08,MITE}, brane-word models \cite{RS99,DBP00,D01,A02,SS03,M05}, scalar fields \cite{SRSS00,S00,COY00,PR03,CST06}, phantom fields \cite{CKW03}, holographic cosmology \cite{KL99,L04,H04,Si07}, or accounting for exotic substances such as Chaplygin gas \cite{BSSS04,BBS03,DBC04}. There are also alternatives which replace dark energy with an inhomogeneity of a radius of approximately one Gpc \cite{C00,INN02,AAG06,AM06,EM07,B08,GBH08,CFL08,BW09,CBKH09} 
(for a review and explicit examples see Bolejko et al. 2009)
--- these, exceptionally, do not require any new physics. 
In most of these alternatives, an effective dark energy 
equation of state evolves with time.
Therefore, from the observational point of view, the most important discovery would be to detect any possible changes in dark energy over the period of evolution of the Universe or to prove beyond doubt that the cosmological constant is indeed constant. 

The evolution of dark energy manifests itself in its equation of state.
For the perfect fluid, the conservation
equations $T^{\alpha \beta}{}_{;\beta} = 0$ imply
\cite{E71}
\[ \frac{\partial {\rho}_{DE}}{\partial t} + \Theta (\rho_{DE} +p_{DE}) =0, \]
where $\rho$ is energy density, $p$ is pressure, and 
$\Theta$ is the expansion parameter 
(in the Friedmann models $\Theta = 3 H$; $H$ is the Hubble parameter).  
Therefore, the only case when dark energy does not evolve is when 
$\rho_{DE} = -p_{DE}$. Thus,  if only 
$w \equiv p_{DE}/\rho_{DE}$ differs from $-1$ then  dark energy 
evolves.

However, as long as dark energy is not directly detected
any measurement of $w$ is always indirect and relies on the assumption
that some relations hold.
In most cases, it is assumed that the distance-redshift relation is as
in the Friedmann model.
This paper shows that in the perturbed universe, even if $\av{\delta\rho} =0$
the distance relation is not the same as in the unperturbed universe.
Therefore, the study of dark energy
within the homogeneous framework may strongly bias the results of these analyses.

Since the distance relation in its general form is a partial differential equation,
to make the reader more aware of some features of differential equations
a simple example with a harmonic oscillator is presented in Sec. \ref{hamosc}.
In Sec. \ref{dist}, the relation for the distance in the general inhomogeneous case is presented and an analogy with the harmonic oscillator solution is discussed. 
Sec. \ref{evde} discusses the consequences of evolving
dark energy.
In Sec. \ref{futobs}, an example of the Swiss-Cheese
model is presented to shown that small-scale inhomogeneities if not analysed properly can  be mistaken for 
an evolving dark energy.

\section{Harmonic oscillator}\label{hamosc}

\begin{figure}
\includegraphics[scale=0.7]{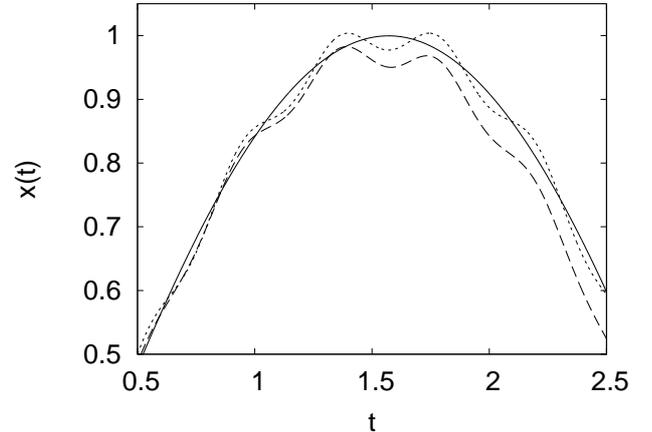}
\caption{The solutions of eqs. (\ref{uho}) -- (\ref{pho2}).
Solid  line: the unperturbed solution, equation (\ref{uho});
dotted line: the solution of equation (\ref{pho1}); dashed line the solution of
equation (\ref{pho2}).}
\label{fig1}
\end{figure}

We now discuss
some basic features of the harmonic oscillator. Although this material is most probably well know,
it is instructive to review it here, as 
there is a important relation to cosmology,
namely to the distance relation.
The evolution of the harmonic oscillator is given by

\begin{equation}
 \ddot{x} + \omega_0^2 x = 0.
\label{uho}
\end{equation}
The solution is a combination of $\sin \omega_0 t$ and $\cos \omega_0 t$. We refer to this solution
as the {\em unperturbed solution}.
If some additional force is present (for simplicity we assume that 
the force is sinusoidal) then

\begin{equation}
\ddot{x} + \omega_0^2 x - F_0 \sin \omega t = 0. 
\label{pho1}
\end{equation}
The solution of this equation is a sum of the 
unperturbed solution and the steady-state solution given by

\[x_s(t) = \frac{F_0}{\omega_0 - \omega} \sin \omega t. \]
As seen, in this case, the driving force introduces oscillations
of frequency $\omega$ around the unperturbed solution.
However, if we depart from this highly idealistic scenario
the solution changes. We assume that $F$ is proportional to $x$ ($F=xF_0 \sin \omega t$), then

\begin{equation}
\ddot{x} + x( \omega_0^2   - F_0 \sin \omega t) = 0.
\label{pho2}
\end{equation}
In this case, the solution does not oscillate around the unperturbed solution.
This is presented in Fig. \ref{fig1}, which presents
the solutions of Eqs. (\ref{uho}) -- (\ref{pho2}).
The parameters used here are
$\omega_0 = 1$, $\omega = 15$, and $F_0=5$.
The unperturbed solution, Eq. (\ref{uho}), is given by the solid line.
The dotted line presents the solution of Eq. (\ref{pho1}) -- as seen, it 
oscillates around the unperturbed solution. The solution of Eq.
(\ref{pho2}) is given by the dashed line: it does not oscillate
around the unperturbed solutions, even though the average of $F$ over
the period vanishes, i.e. $\av{F}_t = 0$.

The above conclusion is rather obvious, although we show there is a close
relation between the above example and the cosmology. We demonstrate that
the distance in the perturbed universe (even if $\av{\delta \rho} = 0$) does not oscillate around
the homogeneous value. This has a strong implication on the analysis of cosmological
observations.

\section{Distance in the inhomogeneous universe}\label{dist}

The Sachs equation for the angular diameter distance is \cite{S61}

\begin{equation}
\frac{{\rm d^2} D_A}{{\rm d} s^2}  = - (
|\sigma|^2 +  \frac{1}{2} R_{\alpha \beta} k^{\alpha} k^{\beta}) D_A,
\label{darel}
\end{equation}
where $D_A$ is the angular diameter distance, 
$\sigma$ is the shear and vanishes in the Friedmann limit,
$R_{\alpha \beta}$ is the Ricci tensor and 
$R_{\alpha \beta} k^\alpha k^\beta = T_{\alpha \beta} k^\alpha k^\beta$.
In the comoving and synchronous coordinates, for pressure-less matter
$R_{\alpha \beta} k^\alpha k^\beta = \kappa \rho k^0 k^0$ ($\kappa=8\pi$ and $G=c =1$).
Since the above relation depends linearly on $\rho$ one could naively presume that since $\av{\delta \rho} = 0$ then the average distance should be the same
as in the unperturbed case, i.e. as in the Friedmann model.

We consider a simple example that shows otherwise: we assume that $\sigma=0$ and $R_{\alpha \beta} k^{\alpha} k^{\beta}$
is constant\footnote{For the homogeneous case, we have $R_{\alpha \beta} k^{\alpha} k^{\beta} \sim a^{-5}$,
so for small distance it changes relatively slowly. Even at distance around 1 Gpc, for the standard cosmological model, $a(t) = 0.8 a_0$ (where $a_0$ is the scale factor at the current instant).}.
For the homogeneous model Eq. (\ref{darel}) has the same form 
as the unperturbed harmonic oscillator in Eq. (\ref{uho}).
However, if we add density perturbations, i.e.
\[ \rho = \rho_0 + \delta \rho\]
then even if $\av{\delta \rho} = 0$ say 
$\delta \rho \sim \sin \omega x$, the
corresponding harmonic oscillator is no longer 
given by Eq. (\ref{pho1}) but by Eq. (\ref{pho2}).
Hence the solution to Eq. (\ref{darel}) for the perturbed universe
does not oscillate about the unperturbed solution,
even if $\av{\delta{\rho}} = 0$.

{\it How does this relate to the real Universe? Is the effect large?}
  
\begin{figure}
\includegraphics[scale=0.7]{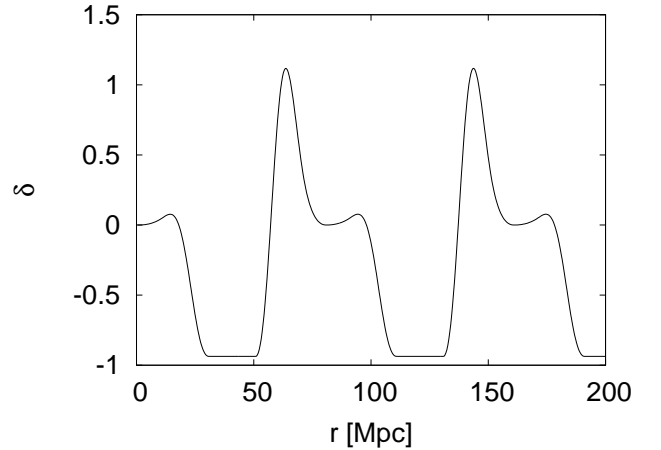}
\caption{Density fluctuations along the line of sight within the lensing and
Swiss Cheese models.}
 \label{fig2}
\end{figure}

The real Universe on small scales is highly inhomogeneous
with cosmic structures such as galaxy clusters, superclusters, and voids.
Cosmic voids are of size 20-50 Mpc \cite{Hoy,Jon}, thus
to test the effect of inhomogeneities on the distance-redshift relations
we consider models with similar scales of inhomogeneity. 
Two types of models are considered: the non-linear Swiss-Cheese model
and the linear perturbative framework. 
The density profile along the line of sight, for both models,
is presented in Fig. \ref{fig2}. 
We consider three  different methods for calculating the distance:

\begin{enumerate}
\item Fully non-linear model:

The distance will be calculated directly from Eq. (\ref{darel}).
The evolution of matter and null geodesics will be calculated 
within the Szekeres Swiss-Cheese model. The details are presented in the Appendix.

\item
Lensing approximation:

Within the linear perturbative scheme the distance is
\begin{equation}
 D_{A}(z)= \bar{D}_{A} ( 1 + \delta_D),
\label{dlen}
\end{equation}
where $\bar{D}_{A}$ is the angular-diameter distance in the homogeneous universe (see Eq. (\ref{dl})). 
The most general form for $\delta_D$ was derived by 
Pyne \& Birkinshaw (2004), Bonvin, Durrer \&  Gasparini (2006), and Hui \& Greene (2006).
Excluding the contribution from the motion of the observer and source, and taking the leading term, $\delta_D$ reduces to

\begin{equation}\label{dDBa}
\delta_D =
- \int\limits_0^{\chi_e} {\rm d} \chi
\frac{ \chi_e - \chi}{\chi_e} \chi \nabla^2 \phi(\chi),
\end{equation}
where $\chi$ is the comoving coordinate $d \chi = dz/H(z)$ and 
$\phi$ is the gravitational potential that
relates to the density perturbations $\delta \rho$ via the  Poisson equation
$ \nabla^2 \phi = \frac{4 \pi G}{c^2} a^2 \delta \rho$.
Equation (\ref{dDBa}) is equivalent to the 
convergence in the lensing approximation and is known as the Born approximation.

The present-day density fluctuations along the line of sight are exactly the same as in the
Szekeres Swiss-Cheese model mentioned above. This allows us
to see how large higher-order corrections are.
The evolution of $\delta$ is calculated using the linear approximations  (Peebles 1980)
\begin{equation}\label{deltaev}
\ddot{\delta} + 2 \frac{\dot{a}}{a} \delta = \frac{4 \pi G}{c^2} \delta \rho. 
\end{equation}

\item Standard, homogeneous, Friedmann model:

The distance  within the homogeneous Friedmann model is

\begin{equation}
 \bar{D}_{A}(z)=\frac{c}{H_0 (1+z)\sqrt{-\Omega_k}}\sin{\left(
\sqrt{-\Omega_k}\int_0^z{\mathrm{d}z'\frac{H_0}{H(z')}}\right)},
\label{dl}
\end{equation}
and
\begin{eqnarray}
&& H^2 =H_0^2\biggl\{\Omega_{m} (1+z)^3+\Omega_{
k}(1+z)^2 + \Omega_{DE} \nonumber \\
&& \times \exp{\left[3\int_0^z
\frac{1+{w(z')}}{1+z'}\mathrm{d}z'\right]}\biggr\},
\label{hz}
\end{eqnarray}
where $w(z)$ is the dark energy equation of state,
and is parametrized as
\begin{equation}
 w(z) = w_0 + w_1 \frac{z}{1+z}.
\label{wzdef}
\end{equation}

\end{enumerate}

\begin{figure}
\includegraphics[scale=0.7]{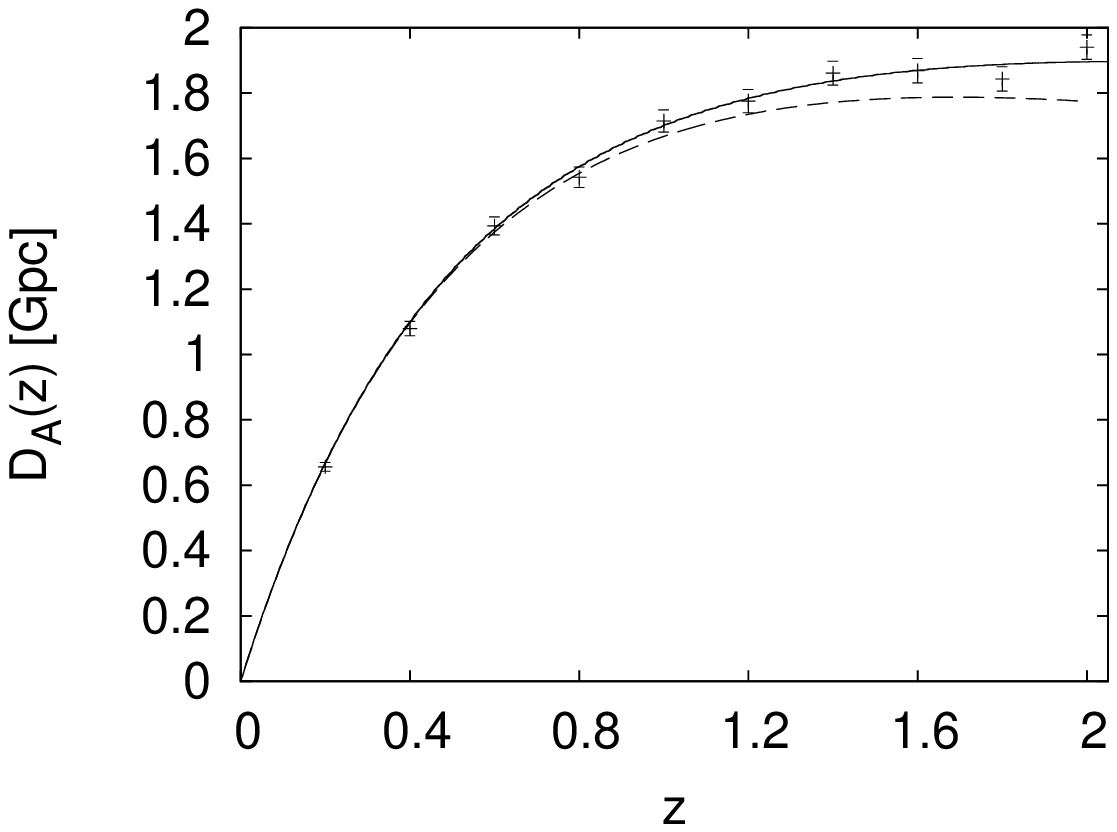}
\includegraphics[scale=0.7]{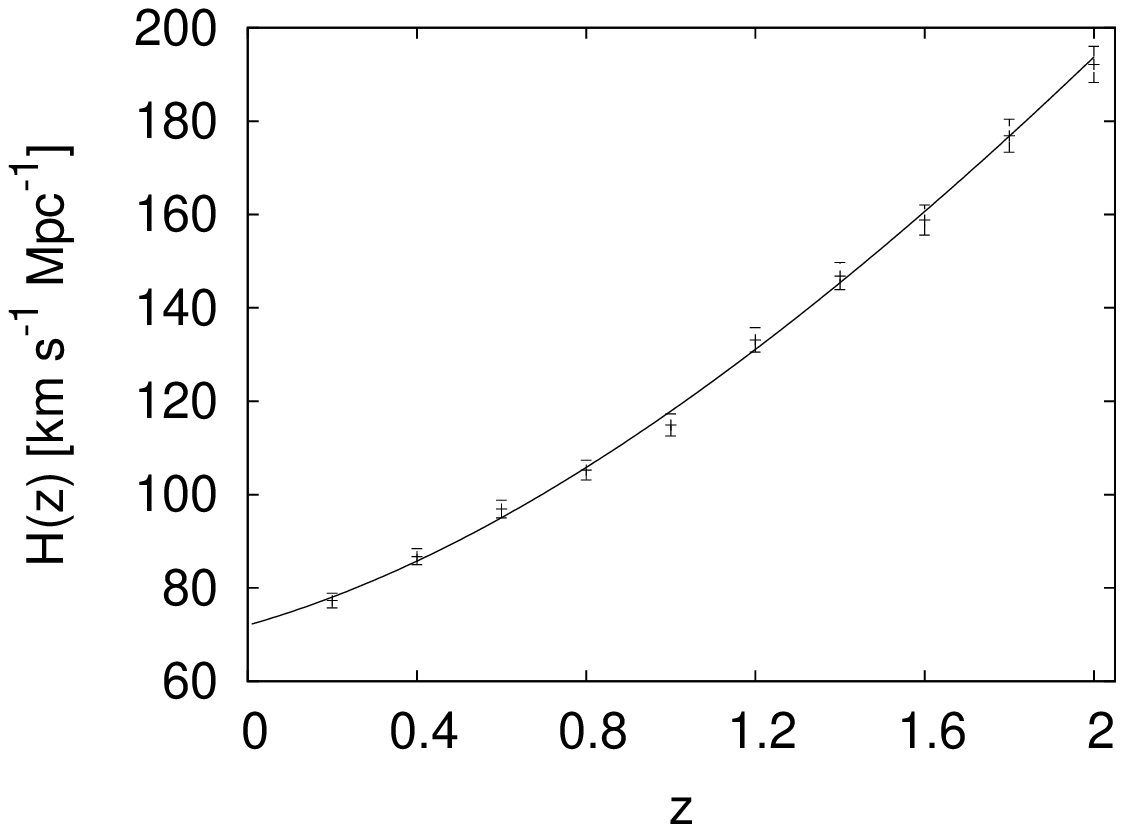}
\caption{The Szekeres Swiss-Cheese model (solid line), the $\Lambda$CDM model (dashed line), and future observations with 2\% precision scattered around the Swiss-Cheese model.
Upper panel presents $D_A(z)$ and lower panel presents the $H(z)$ in the Swiss-Cheese and $\Lambda$CDM model. Since the average $H(z)$ in the Swiss-Cheese model is almost the same as in the background model (see (\ref{avhb})), there is no visible difference between these two models.}
 \label{fig3}
\end{figure}

\section{Evolution of dark energy }\label{evde}

If dark energy is some kind of dynamical field, then its energy density evolves.
Unfortunately, evolving dark energy cannot be modelled using the Szekeres model.
The Szekeres model is the solution of the Einstein equations with a dust source + the cosmological constant, and hence cannot be used to model systems with evolving pressure.
However, evolving dark energy can be modelled using the lensing approximation,
but only in the case when dark energy is homogeneous.
This is because Eq. (\ref{dlen}) assumes that the expansion
rate, which enters via $\bar{D}_A$, behaves as in the homogeneous model.

When dark energy is a dynamical fluid, then two cases need to be considered:
a clustering dark energy and non-clustering dark energy.
If dark energy behaves as a relativistic fluid (like radiation) then its perturbations are described only by decaying modes (Mukhanov 2005).
In this case, dark energy should remain almost homogeneous.

However, if dark energy clusters (like matter) then its preset-day distribution 
can be as inhomogeneous as matter distribution.
Then the lensing approximation [i.e. Eq. (\ref{dDBa})] cannot be applied.
This is because when dark energy is inhomogeneous then the average of the expansion rate
will not be the same as the expansion rate of the background model, i.e. $\av{H} \ne H_0$.
This is a consequence of gradients in the pressure.
If dark energy is inhomogeneous and pressure gradients are non-negligible then 
the $g_{00}$ component of the metric must be position dependent. We denote this by $\alpha$, i.e. $g_{00} = \alpha^2(t,r)$
and for simplicity consider the spherical case. Then (Bolejko \& Lasky 2008)

\begin{equation}
\frac{\alpha'}{\alpha} = -\frac{p'}{\epsilon + p},
\end{equation}
where $\epsilon = \rho_{de} + \rho_{mat}$, $\rho_{de}$ is the energy density of dark energy, and 
$\rho_{mat}$ is the energy density of matter. To  estimate $\alpha$ we assume 
$w_{de} \approx -1$ which implies that $\epsilon + p \approx \rho_{mat}$ and $-p'\approx \rho_{de}'$. 
We also assume 
that the  distribution  of dark energy closely follows the matter distribution, and that
$\rho_{de} \approx 3 \rho_{mat}$ which imply that
\[ \alpha \approx \alpha_o \left( \frac{\rho}{\rho_o} \right)^3 \approx 1+3\delta, \]
where the subscript $mat$ has been suppressed  and the subscript `o' denotes  the `observer's position', i.e. $\alpha_o = \alpha(r=0) \equiv 1, \rho_o = \rho_{mat}(r=0)$. 
The above approximation shows that inhomogeneities in dark energy will have a non-negligible effect
on the dynamics of the Universe. Also, since $\alpha \ne 1 $ the result
of averaging is different than when $\alpha =1$ (i.e. when the pressure gradients are negligible).
For example, Eq. (\ref{avhb}) becomes

\[ \av{H} \approx \frac{4\pi}{V}   \int\limits_0^{r}{\rm d}  {r} (R^2 \dot{R})' \alpha 
\approx \frac{4\pi}{V}   \int\limits_0^{r}{\rm d}  {r} (R^2 \dot{R})' (1+3\delta).
\]
Since $\delta$ is non-zero, $\alpha$ deviates from unity and therefore
\[ \av{H} \ne \frac{\dot{R}}{R} = H_0. \]
This feature was also observed within the second-order perturbative scheme.
Clarkson, Ananda \& Larena (2009) showed that the average quantities (such as the expansion rate) 
within the second-order perturbative framework are not the same as in the unperturbed background, in other words $\av{H} \ne H_0$.
Thus, if dark energy can cluster and is as inhomogeneous as the present-day
matter distribution, then not only the distance fails to oscillate about the unperturbed value but also the expansion rate.
Therefore, in the next section when studying evolving dark energy it is implicitly assumed that dark energy is homogeneous.

\section{Future observations and dark energy}\label{futobs}

\begin{figure*}
\includegraphics[scale=0.47]{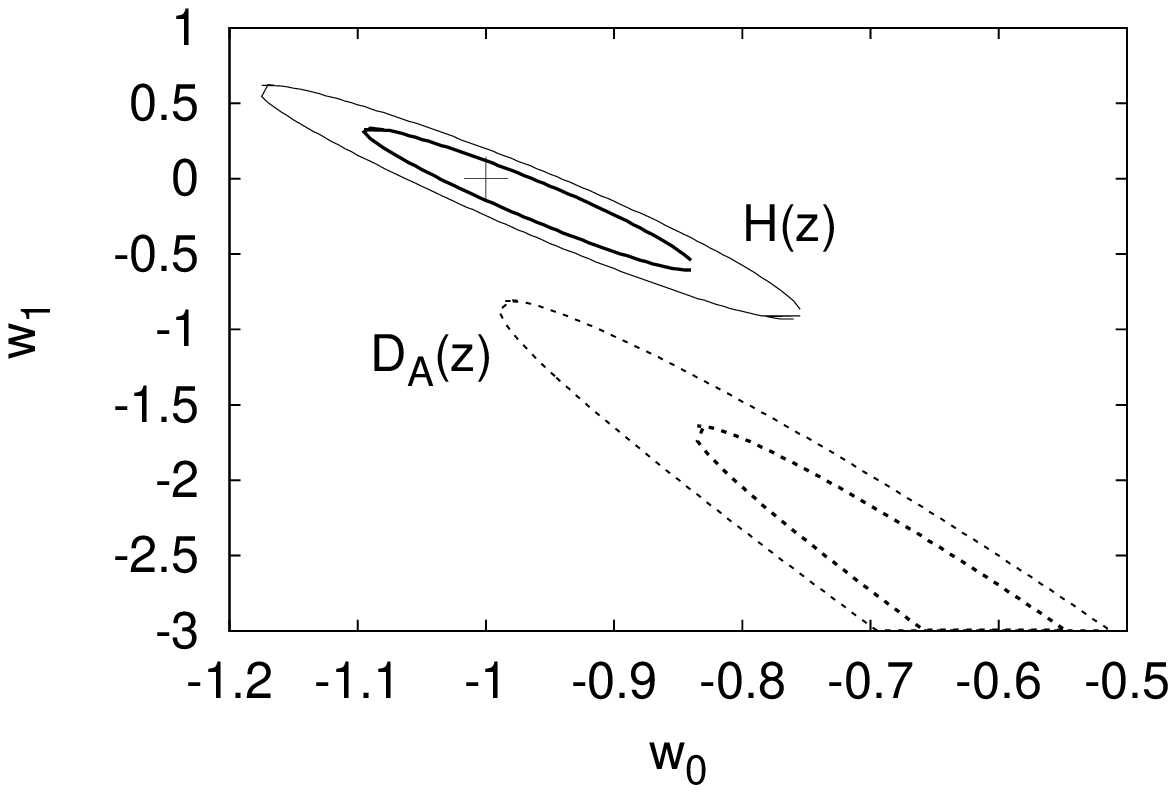}
\includegraphics[scale=0.47]{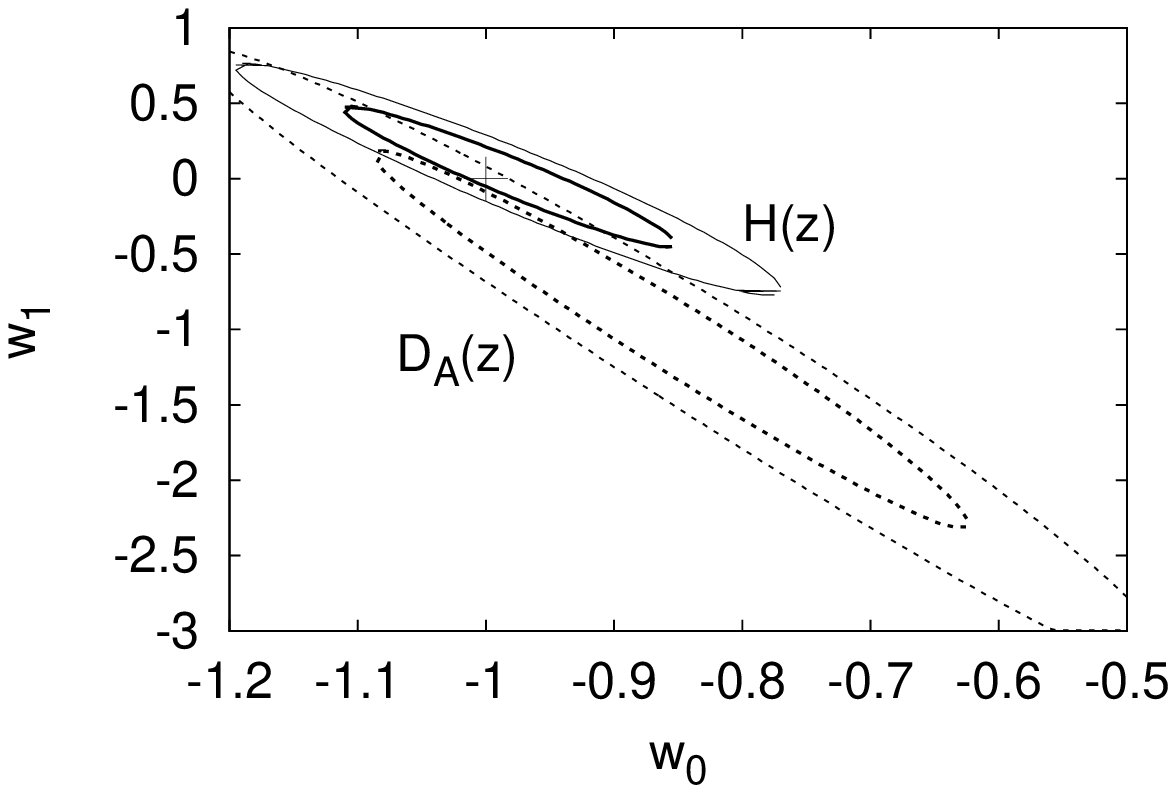}
 \includegraphics[scale=0.47]{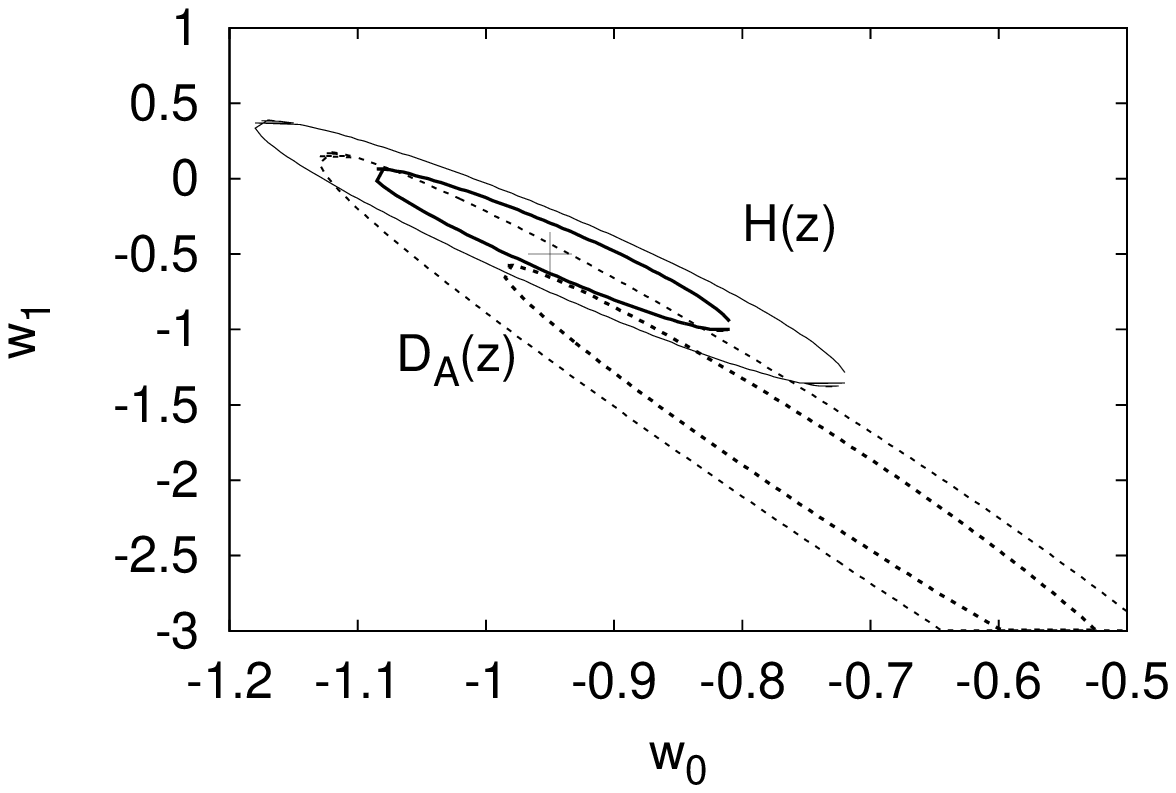}
\caption{68\% and 95\% CL for the parameters $w_0$ and $w_1$ inferred from future (mock) observations.
Solid lines present the constraints from $H(z)$ data, dotted line from 
$D_A(z)$ data, the cross marks the true value of $w_0$ and $w_1$ used to generate the data. The misalignment of distance-constraints is a result
of not taking into account inhomogeneities.
{\em Left}: The results of analysis within the non-linear Swiss-Cheese model ($w_0 = -1$ and $w_1 = 0$).
{\em Middle}: The results of analysis within the linear approximation scheme 
($w_0 = -1$ and $w_1 = 0$). 
{\em Right}:  The results of analysis within the linear approximation scheme 
($w_0 = -0.95$ and $w_1 = -0.5$).}
 \label{fig4a}
\end{figure*}

To test the effects of inhomogeneities 
on the results of the analysis of cosmological observations, we consider the following example.
We assume that we have data from future, very precise measurements of 
$D_A$ and $H$ -- say ten measurements of the baryon acoustic oscillations (BAO) at redshifts $0.2, 0.4, 0.6, ...,1.8, 2$, each measurement having 2\% errors.
We then take the Szekeres model and the lensing approximation, and in each
case generate mock observations (ten BAO data points).
An example of these data is presented in Fig. \ref{fig3},
where generated observations are scattered around the Szekeres model. 
We later analyse these mock observations 
using the standard methods, i.e. 
to find the equation of state of dark energy  
($w_0$ and $w_1$) we analyse the observations by fitting
Eqs. (\ref{dl}) and (\ref{hz}) to the data.
Two scenarios are considered: non-evolving dark energy
(i.e. the cosmological constant) and an evolving one.

The results are presented in Fig. \ref{fig4a}.
The case of non-evolving dark energy is presented in the left and middle panels.
As seen if the data is analysed only using the distance
measurements, the inferred equation of state of dark energy 
differs from $w=-1$. This is because the inhomogeneities
are not taken into account in the distance-redshift Eq. (\ref{dl}).
On the other hand, the analysis of the $H(z)$ data 
agrees  with the true equation of state of dark energy.
This is because, within the Swiss-Cheese model
and within the lensing approximation, the average expansion
rate coincides with the unperturbed value.
However, as pointed out in Sec. \ref{evde}, when dark energy is not
homogeneous, pressure gradients imply that $\av{H} \ne H_0$.
In this case, if pressure gradients are large and not taken into account then 
the inferred value from $H(z)$ data may also differ from the true one.

The evolving dark energy case is presented in the right panel of Fig. \ref{fig4a}.
Only the lensing approximation is considered here, as the Szekeres model
cannot be used to describe the evolution of dark energy.
As seen, the results are very similar to the case of non-evolving dark energy.

The above examples show the importance of inhomogeneities,
as not taking them into account can lead to large systematics,
particularly when measuring the dark energy equation of state.
Evolution of dark energy is a subtle effect, hence  even small
deviations of $D_A(z)$ from the standard case might introduce significant changes in the interpretation of cosmological observations.
Another important result is that non-linear effects are important.
As seen by comparing  the left and middle panels
of Fig. \ref{fig4a}, the non-linear corrections cannot be ignored.

\section{Conclusions}

{\it How do inhomogeneities that are present and observed in the Universe
influence the distance-redshift relation?}

\noindent This question has been frequently addressed  since
the papers by Kristian \& Sachs (1966) and Dyer \& Roeder (1972) 
(see also \cite{SR} and references therein).
However, it is still common among cosmologists to disregard the 
effect of inhomogeneities.
The common reasoning is based on the following two premises:
1) even if density variations are large the fluctuations of the
gravitational potential are small and therefore the perturbation
scheme can be applied, and 2) since perturbations
are Gaussian they vanish after averaging and therefore
they should have little impact on observations.

However, as shown in this paper the distance-redshift relation {\em does} depend on density 
fluctuations (not on gravitational potential), and secondly,
even if perturbations vanish after averaging they do modify the distance-redshift
relation and the final result deviates from the homogeneous one.
Even within the perturbative scheme, it is apparent that 
inhomogeneities do affect the distance-redshift relation \cite{BDG2006,VFW07,GK09}.
Since the evolution of dark energy is a subtle effect, inhomogeneities cannot be ignored.

The main aim of this paper was therefore to show the importance of inhomogeneous models in the studies of dark energy. Ignoring inhomogeneities can have serious consequences since even small-scale inhomogeneities do affect the observations, especially when it comes to the equation of state of dark energy.
This paper uses the explicit example to show that if cosmological observations are analysed within the framework of homogeneous models then matter inhomogeneities might be mistaken for evolving dark energy. This is a consequence of Eq. (\ref{darel}), which is a differential equation for the angular diameter distance. As in the case of non-idealistic harmonic oscillator, the perturbed solution does not oscillate around the unperturbed value. As trivial as it sounds, this statement must be emphasized  because a large number of cosmologists seems to believe otherwise.

The results of this paper imply that systematics due to inhomogeneities can distort our analysis,
and may lead to a conclusion that dark energy evolves  when in fact it is constant (or vice versa).
Since the robust distinction between the evolution and non-evolution of dark energy is the first step 
to understanding its nature, a proper handling of systematics is essential.
 Without taking into account all systematics, the precision cosmology will not be an accurate cosmology (Peebles 2010).

\appendix

\section{Szekeres Swiss-Cheese model}\label{scmA}

The metric of the Szekeres model has the form

\begin{equation}
{\rm d} s^2 =   {\rm d} t^2 - \frac{(R' - R {  E}'/ {  E})^2}
{1 - k} {\rm d} r^2 - \frac{R^2}{E^2} ({\rm d} x^2 + {\rm d} y^2),
\label{ds2}
 \end{equation}
where
\begin{equation}
{  E} = \frac{S}{2} \left[ \left( \frac{x-P}{S} \right)^2
+ \left( \frac{y-Q}{S} \right)^2 +1 \right],
\end{equation}
and the prime denotes the partial derivative with respect to $r$,
$R' = \partial R / \partial r$.
The Einstein equations for the dust are

\begin{equation}
\dot{R}^2 = \frac{2M}{R} - k + \frac{1}{3} \Lambda
R^2, \label{vel}
\end{equation}

\begin{equation}
\kappa \rho = 
 \frac{2M' - 6 M {  E}'/{  E}}{R^2 ( R' - R {  E}'/{  E})}, \label{rho}
\end{equation}
where the dot denotes the partial derivative with respect to $t$,
$\dot{R} = \partial R / \partial t$, $\Lambda$ is the cosmological constant, and 
$M$ is an arbitrary function of $r$. The shear is
\begin{equation}\label{shear}
\sigma^{\alpha}{}_{\beta} = \frac{1}{3} \left( \frac{\dot{R}' - \dot{R}
R'/ R} {R' - R {E}'/{ E}} \right) {\rm diag}
(0,2,-1,-1).
\end{equation}
The bang time function follows from (\ref{vel})
\begin{equation}\label{tbf}
\int\limits_0^{\Phi}\frac{{\rm d} \widetilde{\Phi}}{\sqrt{- k + 2M /
\widetilde{\Phi} + \frac 1 3 \Lambda \widetilde{\Phi}^2}} = t - t_B(r).
\end{equation}
The Weyl curvature decomposed into its electric and magnetic part is
\begin{eqnarray}\label{Weyl}
&& E^{\alpha}{}_{\beta} = 
\frac{M(3 R' - R M'/M)}{3 R^3 ( R' - R E' / E)}
{\rm diag} (0,2,-1,-1), \nonumber \\
&& H^{\alpha}{}_{\beta} =  0.
\end{eqnarray}
The construction of the Szekeres Swiss-Cheese model closely follows \cite{BC10} and the reader is referred
there for more details. The functions that define the Szekeres model were chosen to be of the form

\begin{equation}\label{Mm1}
M = M_0 +  \left\{ \begin{array}{ll}
M_1 \ell^3 & {\rm ~for~} \ell \leqslant x_a, \\
M_2 \exp \left[ -  \left( \frac{ \ell - 2x_a}{x_a} \right)^2 \right] & {\rm ~for~} x_a \leqslant  \ell \leqslant 3x_a \\ 
-M_1 (\ell - 4x_a)^3 & {\rm ~for~} 3x_a \leqslant  \ell \leqslant  4x_a,  \\
0 &  {\rm ~for~} \ell \geqslant 4x_a, 
\end{array} \right.
\end{equation}
where $\ell = r$/kpc, $M_0 = (4 \pi G /3c^2) \rho_{b} \ell^3$,
$\rho_{b} =  \Omega_m \frac{3H_0^2}{8 \pi G}$,
$\Omega_m = 0.25$, $H_0=72$ km s$^{-1}$ Mpc$^{-1}$,
$x_a= 10^4$,
$M_1 = x_a^{-3} M_2 {\rm e}^{-1.5}$, $M_2= -7 \times 10^{11}$ kpc, 

The bang time function is $t_B = 0$, which means that the
age of the universe is the same everywhere. The function $k$ then follows from
Eq. (\ref{tbf}). And the functions $S,P,$ and $Q$ are
\begin{eqnarray}
&& S = (5 \times 10^3 + \ell)^{\pm 0.78} \\
&& P = 1= x_0, \\
&& Q = 1 =y_0.
\end{eqnarray}
where $+$ is for propagation from the origin
[${  E}'/{  E} = 0.78/(5 \times 10^3 + \ell)$], and $-$ towards the origin
[${  E}'/{  E} = -0.78/(5 \times 10^3 + \ell)$].
As can be seen from Eqs. (\ref{rho})--(\ref{Mm1}),
for $r>40$ Mpc the considered model becomes the homogeneous Friedmann model,
which is in this particular case the $\Lambda$CDM model.
First light propagates towards the center, $E'>0$, and after passing through the origin, $E'$ becomes negative, and so on.
Because the inhomogeneous blocks are matched to the Friedmann model
the average density  is almost the same as in the $\Lambda$CDM model

\begin{eqnarray}
&& \av{\rho} =  \frac{1}{V} \int\limits_0^{r}{\rm d}  {r} \int\limits_{-\infty}^{-\infty} {\rm d}  {p} 
\int\limits_{-\infty}^{-\infty} {\rm d}  {q}
\frac{R^2}{E^2}
\frac{R' - R E'/E}{\sqrt{1-k}} \rho = \nonumber \\
&&  \frac{4 \pi}{V}
\int\limits_0^{r}{\rm d}  {r} 
 \frac{R^2 R'}{\sqrt{1-k}} \rho = \frac{1}{V}
\int\limits_0^{r}{\rm d}  {r} 
 \frac{M'}{\sqrt{1-k}} 
\approx \frac{M_0}{V} 
\end{eqnarray}
The first equality is implied by 
the dipole not contributing to the average \cite{KBCMB}.
The last approximation is exact when curvature vanishes,
although in the considered model  $k < 10^{-5}$, so it is 
quite an accurate approximation.
Finally, the junction conditions (matching the Szekeres to Friedmann model)
imply that both the total mass and volume of the inhomogeneous patch are
the same as in the homogeneous model. Thus, the average
density of the Szekeres Swiss-Cheese model is almost
the same as the background model -- the $\Lambda$CDM model.
This is also true for the Hubble parameter

\begin{eqnarray}
&& \av{H} =  \frac{1}{V} \int\limits_0^{r}{\rm d}  {r} \int\limits_{-\infty}^{-\infty} {\rm d}  {p} 
\int\limits_{-\infty}^{-\infty} {\rm d}  {q}
\frac{R^2}{E^2}
\frac{(R' - R E'/E)^2}{\sqrt{1-k}} \times \nonumber \\
&& \frac{1}{3} \frac{\dot{R'} + 2 \dot{R}{R'}/ R - 3 \dot{R} {E}'/{E}}{R' - R {E}'/{E}}  = \nonumber \\
&&  \frac{4 \pi}{3 V}
\int\limits_0^{r}{\rm d}  {r} 
 \frac{R^2 R'}{\sqrt{1-k}} \left(
 \frac{\dot{R}'}{R'} + 2 \frac{\dot{R}}{R} \right)
\approx \nonumber \\
&&   \frac{4 \pi}{3 V} \int\limits_0^{r}{\rm d}  {r} (R^2 \dot{R})' = \frac{\dot{R}_0}{R_0} = H_0
\label{avhb}
\end{eqnarray}

The redshift is obtained for null geodesic equations. The axially null geodesic (propagation along the axial axis) is
\begin{equation}
 \frac{{\rm d} t}{{\rm d} r} = \pm  \frac{R' - R {  E}'/{  E}}{\sqrt{1 - k}}.
\label{snge}
\end{equation}
and the redshift relation is

\begin{equation}
\ln (1+z) = \pm \frac{1}{c} \int\limits_{r_e}^{r_o} {\rm d} r
\frac{ \dot{R}' - \dot{R} {  E}'/{  E}}{\sqrt{1 - k}},
\label{srf}
\end{equation}
or equivalently
	
\begin{eqnarray}
 \frac{{\rm d} r}{{\rm d} z} = \pm \frac{1}{1+z} \frac{\sqrt{1-k}}{
\dot{R}' - \dot{R} {  E}'/{  E}}, \nonumber \\
 \frac{{\rm d} t}{{\rm d} z} = \frac{1}{1+z} \frac{R' - R {  E}'/{  E}}{
\dot{R}' - \dot{R} {  E}'/{  E}}.
\label{redrel}
\end{eqnarray}

\end{document}